\documentclass[doublecol]{epl2} 
\usepackage{color}
\usepackage{amsmath,amssymb,amsfonts}

\newcommand{\Pe}{\operatorname{Pe}} 
\newcommand{\Pu}{\rho_{\uparrow}\!}
\newcommand{\Pd}{\rho_{\downarrow}\!}

\newcommand{\Pp}{\rho}
\newcommand{\Pm}{\delta\rho}
\newcommand{\Ppdot}{\dot\rho}
\newcommand{\Pmdot}{\dot{\delta\rho}}
\newcommand{\Jp}{J_\rho}
\newcommand{\Jm}{J_{\delta\rho}}

\newcommand{\dd}{\mathrm{d}} 
\newcommand{\kappac}{\kappa_\mathrm{c}}
\newcommand{\derx}{\partial_x}
\newcommand{\Pesmall}{\Pe_\text{small}}
\newcommand{\Pelarge}{\Pe_\text{large}}

\DeclareMathAlphabet\mathbfcal{OMS}{cmsy}{b}{n}

\title{Mechanical pressure and work cycle of confined active Brownian particles}
\shorttitle{Mechanical pressure and work cycle of confined ABPs} 

\author{Paolo Malgaretti\inst{1,2,3} \and Piotr Nowakowski\inst{1,2} \and Holger Stark\inst{4}}
\shortauthor{P. Malgaretti \etal}

\institute{                    
  \inst{1} Max--Planck--Institut f\"{u}r Intelligente Systeme, Heisenbergstr.~3, 70569
Stuttgart,\,Germany\\
  \inst{2} IV.\,Institut f\"ur Theoretische Physik, Universit\"{a}t Stuttgart, Pfaffenwaldring 57, 70569 Stuttgart,\,Germany\\
  \inst{3}  Helmholtz Institut Erlangen--N\"urnberg  for Renewable Energy (IEK--11), Forschungszentrum J\"ulich, F\"urther Str.~248, 90429 N\"urnberg,\,Germany\\
  \inst{4} Institut f\"ur Theoretische Physik, Technische Universit\"at Berlin, Hardenbergstr.~36, 10623 Berlin,\,Germany
}

\abstract{We derive an analytic expression for the mechanical pressure of a generic one--dimensional model of confined active Brownian particles (ABPs)
that is valid for all values of P\'eclet number $\Pe$ and all confining scenarios. Our model reproduces the known scaling of bulk pressure with $\Pe^2$ while in strong confinement pressure scales with $\Pe$. Our analytic results are very well reproduced by simulations of ABPs in 2D. We use the pressure formula to calculate both the work performed by an active engine and its efficiency. In particular, efficiency is maximized for work cycles with finite period and not in the limit of infinitely slow cycles as in thermodynamic engines.}

\begin{document}

\maketitle

\section{Introduction}
The properties and structure formation of active systems are
quite different as compared to their equilibrium counterparts
\cite{Ramaswamy2010,RomanczukSchimansky-Geier2012,MarchettiSimha2013,ElgetiReview,Zoettl2016,Bechinger2016,Gompper2020}.
This
becomes particularly apparent in confinement
\cite{Lee2013,Yang2014,Saintillan2015,Cottin-Bizonne2015,Malgaretti2016,Malgaretti2017,Fily2017,Ostapenko2018,Das2018,Peter2020}. Indeed, active particles accumulate at walls~\cite{Elgeti2013}, interfaces~\cite{Simmchen2017,Malgaretti2018}, 
as well as obstacles~\cite{Tagaki14,Zeitz17}, 
and in denser suspensions they show motility--induced phase separation~\cite{Cates2015}. 
One of the key macroscopic quantities of interest is the mechanical pressure $\Pi$ that active particles exert on confining walls
\cite{Brady2014,Solon2015,Saintillan2015,Wittmann2019}.
Such a quantity is crucial for determining the performance of devices rectifying active motion~\cite{Sokolov2009,Di_Leonardo2010,Kaiser14}, for work cycles that exploit active baths~\cite{Fodor2020}, for invasion of active particles into confining space~\cite{Doostmohammadi2019}, as well as evaporation~\cite{Callegari2019} and wetting~\cite{Nestor2017,Duzgun2018} in active fluids .

Predicting the value of $\Pi$ for active systems is not trivial since, due to the active nature of the particles, pressure is, in general, no longer a thermodynamic state function \cite{Tailleur2015}. It is not even an intensive variable 
since it 
depends explicitly on extensive variables (like number of particles), as we show below.
Several articles~\cite{Brady2014,Solon2015,Saintillan2015} have reported independently the characteristic scaling for the pressure, $\Pi\propto \Pe^2$, 
where the P\'eclet number $\Pe=v_{\text{act}}R/D$ depends on the active velocity $v_\text{act}$, the linear size $R$, and the diffusion coefficient $D$ of the particle. 
This relation for pressure has been derived for semi--infinite systems. 
However, the dynamics of active Brownian particles (ABPs) is very sensitive to the presence of boundaries~\cite{Lee2013,Yang2014,Saintillan2015,Cottin-Bizonne2015,Malgaretti2016,Malgaretti2017,Fily2017,Ostapenko2018,Das2018} 
and it is not obvious that the scaling for the pressure also holds 
for \textit{confined} ABPs. 
Indeed, recent numerical works~\cite{Yang2014,Saintillan2015} have shown that the pressure of strongly confined active 
particles scales as $\Pi\propto \Pe$. Therefore, the scaling of $\Pi$ with $\Pe$ depends on the system size. 
At the moment, a comprehensive relation, 
valid for all confining scenarios, between the pressure and the microscopic parameters (such as active velocity and tumbling rate),
which control the dynamics of ABPs, is still lacking.

In this letter we derive a closed--form expression for the pressure exerted by confined ABPs  
that is valid for all values of $\Pe$ and all confining scenarios.
In order to do so, we consider simple ABPs that only move in one dimension either along the $x$ axis (``up'' state) or against it (``down'' state) and that tumble between both states. 
Furthermore, the ABPs experience a confining soft potential
such that, in the limit of diverging potential strength,
our model retrieves the case of ABPs confined in a box with hard walls. In this standard case, the calculated mechanical pressure displays multiple scalings with $\Pe$. In particular, when particles undergo multiple tumbling events between subsequent collisions with the walls (diffusive regime), pressure scales as $\Pi\propto \Pe^2$ in agreement with Refs.~\cite{Brady2014,Tailleur2015,Saintillan2015}. 
In contrast, for strongly confined active colloids or for very large values of $\Pe$ (as the one attained for dry macroscopic active matter \cite{Moussaid2016,Volpe2016,Saloma2003,Altshuler2005,Zuriguel_sheep_2020,Zuriguel2020}) particles only undergo a few (if at all) tumbling events (ballistic regime) and the pressure scales as $\Pi\propto \Pe$.
Our analytic predictions for the pressure are in very good agreement with results from numerical simulations of ABPs in two dimensions, with the numerical results of  Refs.~\cite{Yang2014,Saintillan2015} and with the expansion approach of Ref.~\cite{Duzgun2018} which exclusively treats the case where pressure scales as $\Pi \sim \Pe^2$.
Thus, despite its simplicity our 
model captures the essence of the dynamics of confined ABPs for all values of $\Pe$. 
Hence, it can be 
used to predict the mechanical 
pressure of ABPs on both the microscopic ($\sim \mu \mathrm{m}$) and macroscopic ($\sim \mathrm{cm}, \mathrm{m}$) scale in all confining 
scenarios. 


We apply the analytical formula for the mechanical pressure to the recently introduced work cycle of active engines~\cite{Fodor2019,Fodor2020,Kroy2020,Holubeck2020}; devices that
exploit the capability of a bath of ABPs to perform directed work (see also a recent review on the topic Ref.~\cite{Fodor_arxiv}). We find that the work is governed by two dimensionless
parameters and that the efficiency of
quasistatic work cycles is optimal for a finite period in contrast to thermodynamic engines.


\section{Model}
The $N$ noninteracting ABPs experience the confining potential 
\begin{align}
    \beta\, U\left(z\right)=\begin{cases}
    f \left(z/L-1\right) & z> L,\\
    0 & -L \leqslant z \leqslant L,\\
    -f \left(z/L+1\right) & z< -L,
    \end{cases}
    \label{eq:txt_pot}
\end{align}
where $L$ is the size of the system (not including the soft walls), $f$ controls the softness of the walls, $\beta=1/\left(k_\mathrm{B} T\right)$, $k_B$ is the Boltzmann constant, and $T$ is the temperature. Within the overdamped regime, the time evolution of the reduced densities for up and down states, ``up'' ($\Pu$) and ``down'' ($\Pd$) states, which we express as functions of the dimensionless position $x=z/L$ and with time in units of $L^2/D$, are governed by
\begin{subequations}\label{eq:smol}
\begin{align}
\dot{\rho}_{\uparrow}\left(x\right)
& =  -\derx J_{\uparrow}-\Gamma
\frac{L^2}{R^2}
\left[\Pu\left(x\right)-\Pd\left(x\right)\right],\\
\dot{\rho}_{\downarrow}\left(x\right)
& =  -\derx J_{\downarrow}+\Gamma 
\frac{L^2}{R^2}
\left[\Pu\left(x\right)-\Pd\left(x\right)\right].
\end{align}
\end{subequations}
(For simplicity we do not denote explicitly the dependence on time.) In Eqs.~\eqref{eq:smol} we have identified the fluxes as
\begin{subequations}\label{eq:fluxes}
\begin{align}
\!\!\!\!J_{\uparrow}\left(x\right) &\! = \! -\!\left[\derx \Pu\left(x\right)-\frac{L}{R}\Pe \Pu\left(x\right)+\Pu\left(x\right)\beta\, \derx U\left(x\right)\right]\!,\!\!\\
\!\!\!\!J_{\downarrow}\left(x\right) &\! = \! -\!\left[\derx \Pd\left(x\right)+\frac{L}{R}\Pe \Pd\left(x\right)+\Pd\left(x\right)\beta\, \derx U\left(x\right)\right]\!,\!\!
\end{align}
\end{subequations}
and we have introduced 
\begin{align}\label{eq:txt_def-Pe-Gamma}
\Pe & =v_{\text{act}}R/D, & \Gamma &= \gamma R^{2}/D,
\end{align}
the particle P\'eclet number $\Pe$ and dimensionless tumbling rate $\Gamma$ defined as
tumbling rate $\gamma$  
times the diffusion time scale $R^2/D$. For later use, we note that $\Pe^2/\Gamma = v_{\text{act}}^2 \gamma^{-1} / D$ 
is the ratio of active to passive diffusion coefficients. We note the ratio $\Pe/\Gamma=v_{\text{act}}/\left(\gamma R\right)$ can be identified with the dimensionless rotational P\'eclet number $\Pe_\mathrm{r}$.

By solving Eqs.~\eqref{eq:smol} in steady state using piecewise solutions in the three regions of $\beta\, U(x)$ (see Sec.~S1 of the Suppl.\ Mat.), we compute the dimensionless mechanical pressure exerted on the right wall (the same results hold for the left wall),
\begin{equation}
   \Pi=
   \int_1^\infty \left[\Pu\left(x\right)+\Pd\left(x\right)\right]f \dd x.
\end{equation}

\section{Pressure} In order to 
study the case of ABPs confined within a box, we take the limit 
of the hard--core potential ($f\to \infty$, see Sec.~S1 of the Suppl.\ Mat.).
In this limit,
$\Pi$ becomes
\begin{equation}
     \Pi_\infty=
     \bar{\rho}\,\frac{R^2}{L^2}\frac{\kappac^3 \cosh \kappac}{\Pe^2\sinh \kappac+2\Gamma \kappac\cosh \kappac}.
    \label{eq:txt_Pr-box}
\end{equation}
Here, $\bar{\rho}=N R/2L$  is the dimensionless number density and 
\begin{equation}
    \kappac=\kappa L=\frac{\sqrt{\Pe^2+2\Gamma}}{R}L \, ,
    \label{eq:txt_kappa}
\end{equation}
where $\kappa$ is the inverse of the effective length that characterizes the exponential decay of the density profile close to the wall\footnote{The exponential decay and the associated decay length (Eq.~\eqref{eq:txt_kappa}) are valid for all values of $\Pe$ and $L$. In particular, for small values of $\Pe$ Eq.~\eqref{eq:txt_kappa} reduces to $\kappa\simeq \Gamma$, in agreement with Ref.~\cite{Elgeti2013}.}.
We remark that $\kappa$ depends solely on microscopic parameters and not on the system size. 
In particular, when $\kappac\gg 1$, Eq.~\eqref{eq:txt_Pr-box} 
is approximated by
\begin{equation}
     \Pi_\infty\simeq\bar{\rho}\,\frac{R^2}{L^2}\frac{\kappac^3}{\Pe^2+2\Gamma \kappac} \, .
    \label{eq:txt_Pr-box-approx}
\end{equation}
The regime $\kappac\gg 1$ is typical for active matter as it occurs whenever either $\Pe\gg R/L$ or $\Gamma\gg R^2/L^2$.\footnote{We remark that if 
 $L/R>10$, $\kappac\gg 1$ provided that $\Pe \geqslant 0.1$ or $\Gamma\geqslant 0.01$.}
The latter means that during passive diffusion across the system, tumbling occurs frequently.
Therefore, in the following we focus on the relevant case of $\kappac\gg 1$.
In Sec.~S2 of the Suppl.~Mat.\ we present mathematical derivation of all limiting regimes discussed in this letter and argue that the results are valid even for moderately large values of $\kappac$.

\begin{figure}
    \centering
    \includegraphics[scale=0.54]{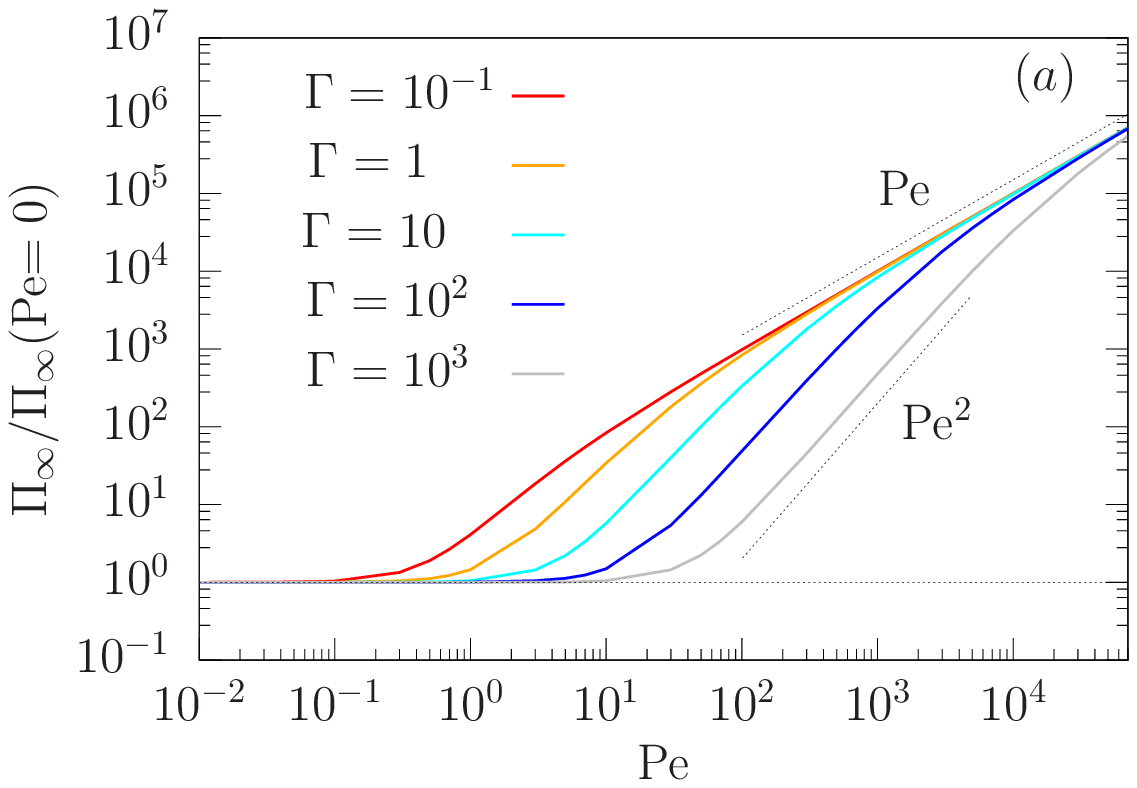}
    \includegraphics[scale=0.54]{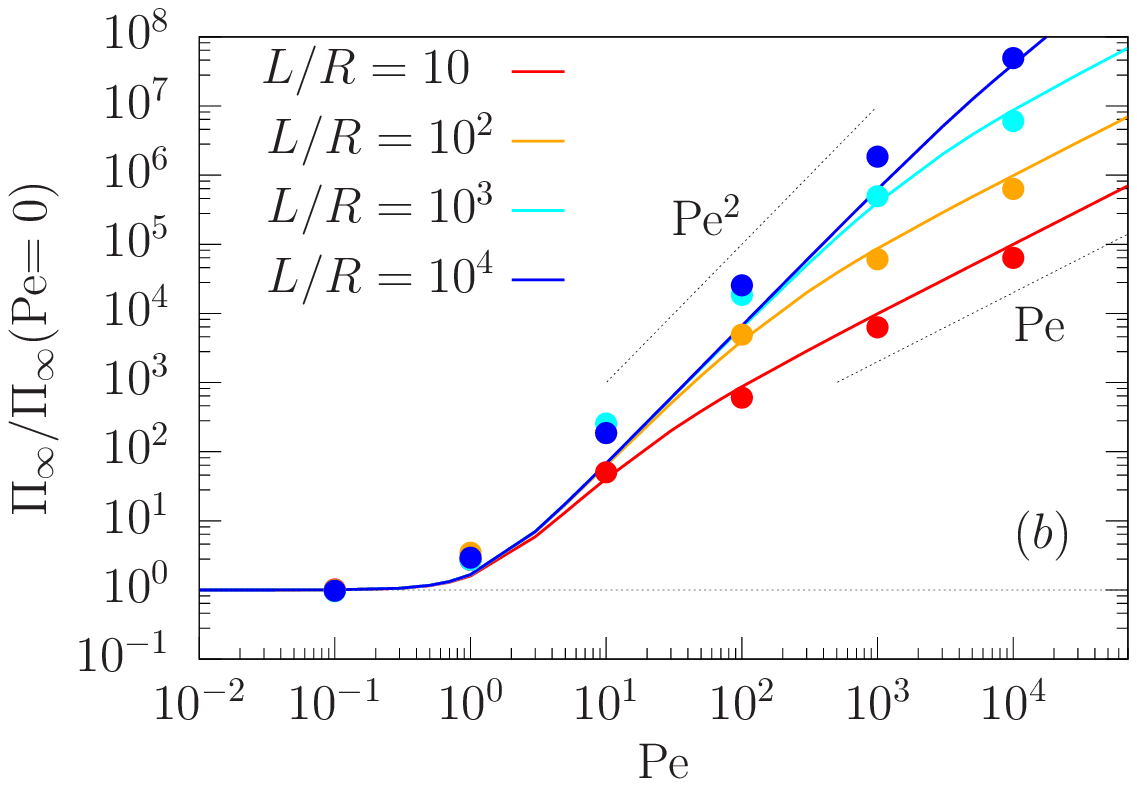}
    \caption{(a): Pressure $\Pi_{\infty}$ as function of $\Pe$ for diverse values of $\Gamma$
    and $L=10 R$. (b): $\Pi_{\infty}$ as function of 
    $\Pe$ for $\Gamma=3/4$ and diverse values of $L/R$. 
    The dots present the results of $2D$ numerical simulations of ABPs characterized by the same value of $\Pe$,  the linear size $L$ and $\Gamma$, and with no fitting parameters (see Sec.~S6 of the Suppl.~Mat.).}
    \label{fig:pressure_model}
\end{figure}

Figures~\ref{fig:pressure_model}(a) and (b) show that $\Pi_\infty$ grows monotonically upon increasing $\Pe$, as expected.
In the limit of small P\'eclet numbers, $\Pe\ll \Pesmall=\sqrt{2\Gamma}$, Eq.~\eqref{eq:txt_Pr-box-approx} gives
\begin{equation}
    \Pi_\infty^\mathrm{small}\simeq \bar{\rho}  \left(1+ \frac{\Pe^2}{2\Gamma} \right)\,,\quad \Pe\ll\Pesmall \, .
    \label{eq:pi_small}
\end{equation}
Thus, for vanishingly small $\Pe$,  $\Pi_\infty$ reduces to its equilibrium value $\Pi^0_\infty = \bar{\rho}$.
We note that 
 the expression in the brackets agrees with the effective temperature
  $T_\mathrm{eff}= T(1+\Pe^2/2\Gamma)$ introduced, for example, in Refs.~\cite{Howse2007,Palacci2010} (see also Sec.~S3 of the Suppl.~Mat.).
We remark that $\Pi^0_\infty$ does not depend explicitly on system size $L$ or particle number $N$, as required 
for an intensive thermodynamic quantity.
For $\Pe\simeq \Pesmall$, activity starts to dominate the pressure. Using Eqs.~\eqref{eq:txt_def-Pe-Gamma},
the condition $\Pe\simeq \Pesmall$ implies
$D_{\mathrm{act}}=v_{\mathrm{act}}^2/\gamma\simeq D$. Thus, the pressure starts to grow with $\Pe^2$
when the active contribution to the total diffusion coefficient $D_{\mathrm{eff}}=D+D_{\mathrm{act}}$ becomes dominant \cite{Howse2007,Palacci2010}.

For large P\'eclet numbers ($\Pe\gg \Pesmall$) we obtain from Eq.~\eqref{eq:txt_Pr-box-approx}:
\begin{subequations}\label{eq:pi_large}
\begin{align}
\label{eq:pi_largeII}	\Pi_\infty^\mathrm{large}&\simeq \bar{\rho} \frac{\Pe^2}{2\Gamma}\,, && \text{for } \Pelarge\gg\Pe\gg \Pesmall,\\
\label{eq:pi_largeI}		\Pi_\infty^\mathrm{large}&\simeq \bar{\rho} \frac{L}{R} \Pe\,, && \text{for }\Pe\gg \Pelarge,\Pesmall,
\end{align}
\end{subequations}
where $\Pelarge=2\Gamma L/R$.
When $\Pe\gg\Pelarge$, pressure $\Pi_\infty$ attains its asymptotic form \eqref{eq:pi_largeI}, $\Pi_\infty \simeq 
N \Pe/2$,  
growing linearly in $\Pe$. In this regime the pressure 
is no longer an intensive variable since it depends explicitly on particle number $N$.

The crossover between the different scalings of $\Pi_\infty$ with $\Pe$ occurs at $\Pe = \Pelarge=2\Gamma L/R$, \textit{i.e.}, 
when the ballistic time $L/v_{\mathrm{act}}$ is comparable to the mean run time $1/\gamma$ between two tumbling events. 
Thus, in the asymptotic regime where
$L/v_{\mathrm{act}}\ll 1 / \gamma$, particles only
undergo a few (if at all) tumbling events between subsequent collisions with the 
walls (ballistic regime).
As a result, they spend the large amount of time at the boundaries.
Therefore, the pressure increases linearly with $\Pe$ and depends explicitly only on particle number
and not on system size. In the opposite case $L/v_{\mathrm{act}}\gg 1 / \gamma$ (\textit{i.e.}, $\Pe\ll\Pelarge$), the particles undergo multiple tumbling events between subsequent collisions with the walls 
(diffusive regime). In this regime, increasing $\Pe$ has a twofold effect:
first, it directly enhances pressure when the particles hit the wall,
and second, it reduces the number of tumbling events between to subsequent collisions at the walls and thereby enhances the density of the particles at the wall. 
This twofold effect explains the quadratic dependence of $\Pi_{\infty}$ on $\Pe$ in Eq.~\eqref{eq:pi_largeII}, as shown in both panels of Fig.~\ref{fig:pressure_model}.

Accordingly, Fig.~\ref{fig:pressure_model}(a) shows that the scaling of the pressure with $\Pe$ changes dramatically upon changing 
the tumbling rate $\Gamma$. At the micrometric scale, this result is crucial for confined bacterial suspensions~\cite{Najafi2018,Seyrich18}, whose 
tumbling rate depends on both the biology of the bacteria as well as on external control parameters\footnote{Typically, $\Gamma\gtrsim 100$~\cite{Najafi2018}. Even upon genetically switching off tumbling, the lower bound is $\Gamma\simeq 1$ due to rotational diffusion. In addition, in chemical gradients $\Gamma$ can vary by a factor of five \cite{Seyrich18}.}.
At the macroscopic scale our result is crucial for determining the pressure of dry active matter, such as small robots~\cite{Volpe2016}, ants~\cite{Saloma2003,Altshuler2005}, sheep~\cite{Zuriguel_sheep_2020}, and humans~\cite{Moussaid2016,Zuriguel2020}, just to mention  a few among others. 

For active colloids, such as Janus particles, $\Gamma$ is controlled
by the rotational diffusion coefficient $D_{\mathrm{rot}}$, which depends on their size and
shape. For spherical particles
$D_{\mathrm{rot}}=\frac{3}{4}\frac{D}{R^2}$, which in 2D equals $\gamma$
so that $\Gamma=3/4$.
For this case, Fig. \ref{fig:pressure_model}(b) 
presents $\Pi_\infty$ versus $\Pe$ for diverse system sizes. 
Interestingly, for typical values of the P\'eclet number ($\Pe\simeq 1-100$) and system sizes ($L\gtrsim 100R$) that have been investigated experimentally~\cite{Junot2017} or numerically~\cite{Winkler2015}, our model predicts $\Pi_\infty\propto \Pe^2$,
in agreement with Ref.~\cite{Brady2014,Tailleur2015,Solon2015,Saintillan2015,Winkler2015,Caprini2018}. 
However, for smaller system sizes $L\simeq 10 R$ or for very large values of 
the P\'eclet number, $\Pe\gg \Pelarge$, the asymptotic behavior $\Pi_\infty\propto \Pe$ is retrieved.
\begin{figure}
    \centering
    \includegraphics[scale=0.54]{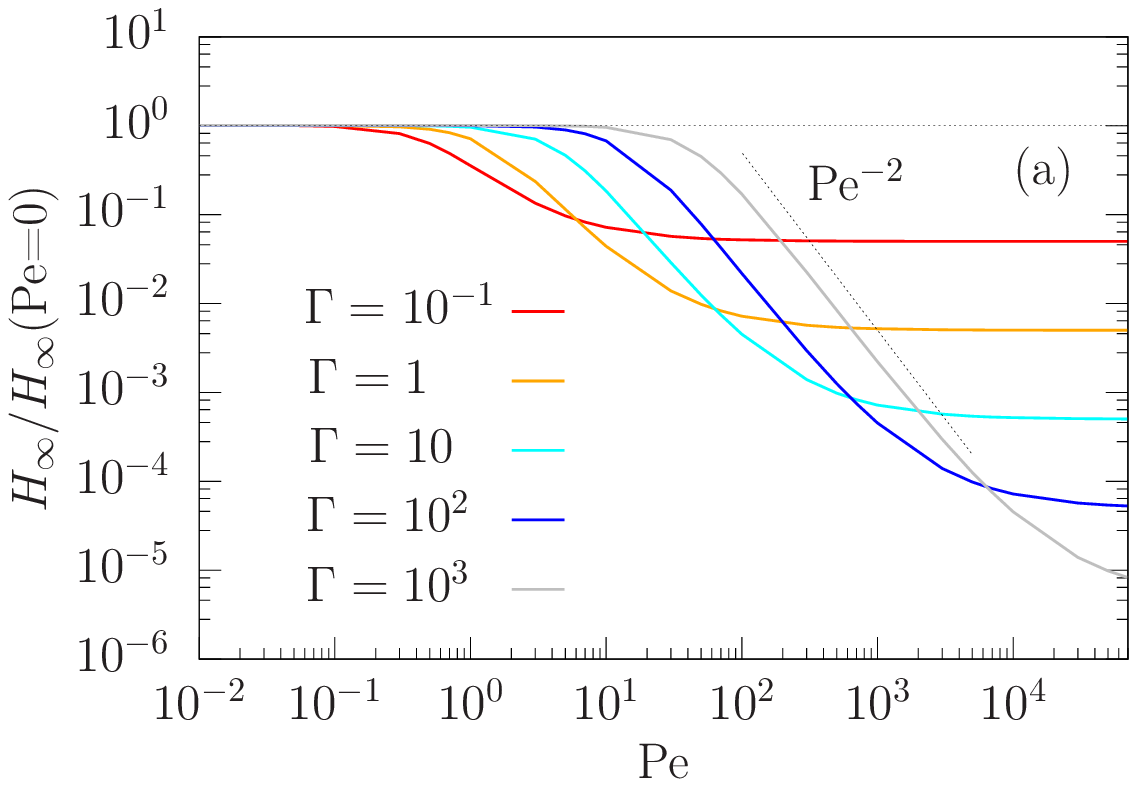}
    \includegraphics[scale=0.54]{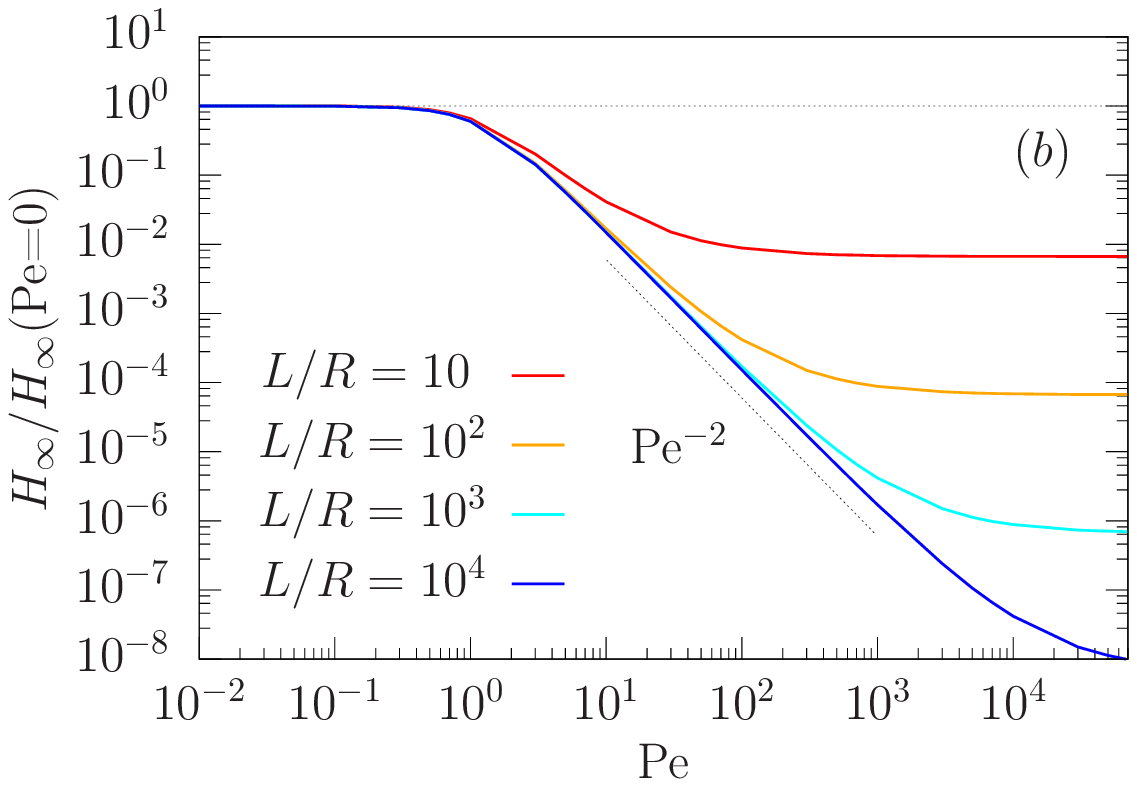}
    \caption{(a): 
    Compressibility
    $H_\infty$ as function of $\Pe$ for diverse values of $\Gamma$ and for $L/R=10$. (b): $H_\infty$ as function of $\Pe$ for diverse values of $L/R$ and for $\Gamma=3/4$. }
    \label{fig:compr}
\end{figure}

In order to check the validity of our expression against more realistic models, we performed 2D simulations of spherical ABPs characterized by  $\Gamma=3/4$, where the particle orientations diffused on the unit circle (see Sec.~S6 of the Suppl.\ Mat.~for more details on the simulations).
The results are included in Fig.~\ref{fig:pressure_model}(b).
Interestingly, without using any fitting parameters, the agreement between the theoretical predictions and the results of the numerical simulations is very good for all values of $\Pe$ and $L$ that we tested. 
Hence, our 
simple two--state model captures the essence of the dynamics of confined ABPs \cite{Cates13}.
\begin{figure*}[t]
    \centering
    \includegraphics[scale=0.6]{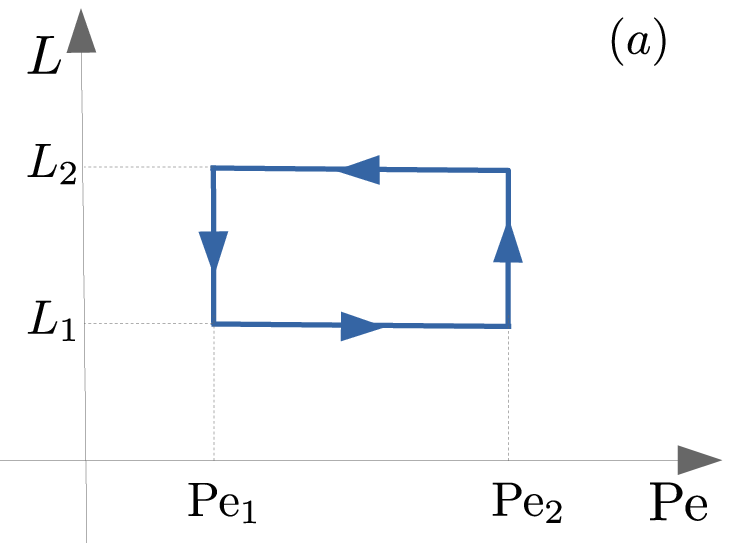}
    \includegraphics[scale=0.45]{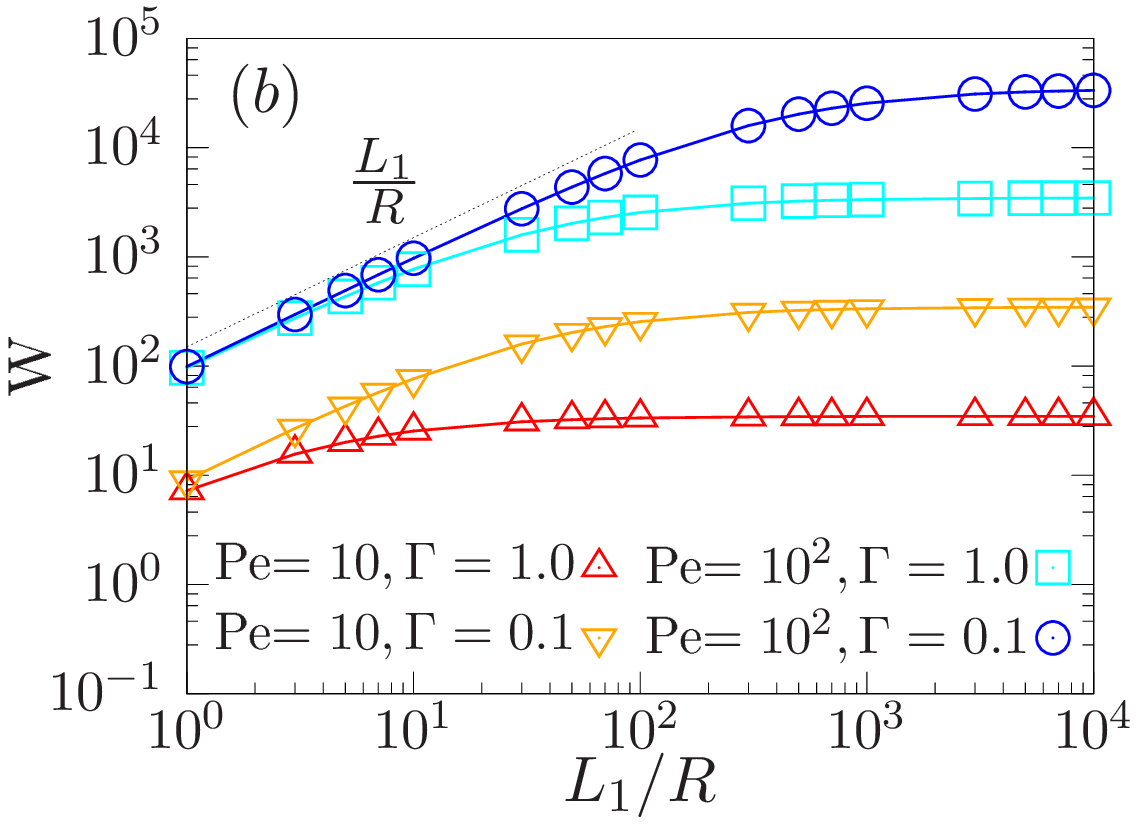}
    \includegraphics[scale=0.45]{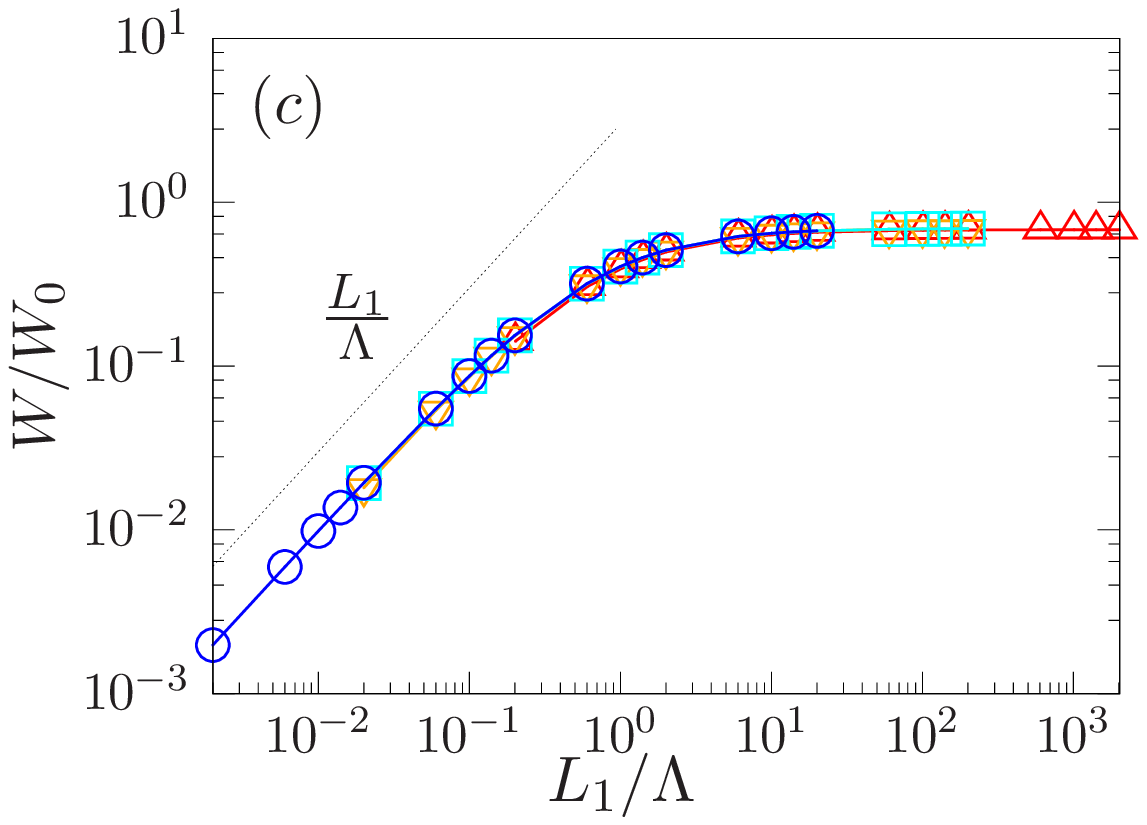}
    \caption{a): Scheme of the work cycle of the active engine in the $\Pe$--$L$ plane.
    b) Work $W$ performed by one cycle of the active engine as function of $L_1$ with $L_2=2L_1$ and $\Pe_1=0$ for diverse values of $\Pe_2 = \Pe$ and $\Gamma$.
    c) Rescaled data of panel (b)}
    \label{fig:cycle}
\end{figure*}
\section{Compressibility} Having an explicit expression for the pressure [cf.~Eq.~\eqref{eq:txt_Pr-box}], we can calculate explicitly the dimensionless 
compressibility 
$H_{\infty}=-(\partial \Pi_{\infty} / \partial L)^{-1}/L$ (see Sec.~S5 of the Suppl.\ Mat.).  For the relevant case of 
$\kappac\gg 1$, it reduces to
\begin{align}\label{eq:compr-approx}
    H_\infty&\simeq \frac{1}{\Pi_\infty}
    \left( 1 +  \frac{\Pe^2}{2\Gamma \kappac}  \right)\,.
\end{align}
Figure \ref{fig:compr}(a) shows $H_\infty$ plotted versus $\Pe$ for
$L=10R$ and diverse values of $\Gamma$.
Similar to the behavior of pressure, upon increasing $\Pe$ beyond $\Pesmall =\sqrt{2\Gamma}$, 
the compressibility starts to decrease 
as $\Pe^{-2}$ due to the prefactor $1/\Pi_{\infty}$ [cf.~Eq.~\eqref{eq:pi_small}] and then, beyond 
$\Pe\simeq\Pelarge= 2\Gamma L / R$,
it reaches the normalized plateau value 
$(N\Gamma L/R)^{-1}$,
\textit{i.e.}, the compressibility depends on extensive variables such as $N$ and $L$.
This is in contrast to equilibrium systems, for which compressibility is an intensive variable whose corrections due to finite size become negligibly small upon increasing system size.
In particular, larger values of $\Gamma$ delay the onset of the decrease of $H_\infty$ and they also lower the plateau value since more tumbling releases pressure generated by the active particles.
Figure~\ref{fig:compr}(b) shows the relevant case of active colloids and that not only pressure $\Pi_\infty$ but also $H_\infty$ retains a dependence on the system size.
 
\section{Active Engine}
We can exploit the exact expression for the pressure to calculate the work performed by the system during the 
periodic work cycle shown in Fig.~\ref{fig:cycle}(a). 
Work is performed by the system only when there is a change in the system size $2L$.  Assuming that these changes are sufficiently slow so that the 
mechanical pressure can
adjust instantaneously, the total dimensionless work per particle along a cycle reads:
\begin{align}
    W\!=\!\!\int\limits_{L_1}^{L_2}\frac{\Pi_\infty\left(L,\Pe_2\right)}{N}
    \frac{2\dd L}{R}\!
    -\!\!\int\limits_{L_1}^{L_2}\frac{\Pi_\infty\left(L,\Pe_1\right)}{N}
    \frac{2\dd L}{R}\,. 
    \label{eq:work}
%
\end{align}
Figure~\ref{fig:cycle}(b) shows that for small system sizes the work performed by the system grows linearly with $L_1$
(for fixed ratio $L_2/L_1$).
In this regime the work is insensitive to $\Gamma$
[blue and cyan curves as well as red and orange curves lie on top of each other in Fig.~\ref{fig:cycle}(b)],
while the overall amount of work depends on $\Pe$.
Upon increasing $L_1$ further, the work $W$ reaches a plateau. 
Here, $W$ increases upon decreasing $\Gamma$ for both values of  $\Pe$, since tumbling reduces the pressure 
of the expanding system.
The dependence of $W$ on $\Gamma$ 
shows that the work performed by the active system explicitly 
depends on the tumbling rate, \textit{i.e.}, on some microscopic time scale. Such a dependence does not occur in passive systems and therefore is a signature of the active nature of the system under study.
In contrast, the dependence of $W$ on $\Pe$
is clear since the active motion of the ABPs generates the force with which they push against the wall.

All these observations can be rationalized by considering the limit $\kappac\gg 1$ in Eq.~\eqref{eq:work} which gives
\begin{align}
    W\simeq W_0\ln\left[\frac{1+L_2/\Lambda}{1+L_1/\Lambda}\right]-\ln\left[\frac{L_2}{L_1}\right] \, .
    \label{eq:work-approx}
\end{align}
Here, we have identified the effective length $\Lambda$ and strength $W_0$ of the work cycle,
 \begin{align}
    \Lambda=\frac{\Pe^2 R}{2\Gamma\sqrt{\Pe^2+2\Gamma}},&\quad 
W_0 = 1 + \frac{\Pe^2}{2\Gamma}\,.
\end{align}
After rescaling work by $W_0$ and system size by $\Lambda$, for $W_0 \gg 1$ all curves from Fig.~\ref{fig:cycle}(b) collapse onto one
master curve, as demonstrated in Fig.~\ref{fig:cycle}(c).
In the regime 
$\Pe \gg \Pesmall=\sqrt{2\Gamma}$
the parameters become 
\begin{align}
    \Lambda \simeq \frac{v_{\mathrm{act}}}{\gamma},&\quad W_0 \simeq \frac{v^2}{D\gamma}=\frac{D_{\mathrm{act}}}{D},
\end{align}
\textit{i.e.}, the threshold length $\Lambda$ is the typical distance traveled by the particle between two 
tumbling events, whereas the work strength $W_0$ is proportional to the ratio of active to passive diffusion coefficients.
In particular, for large systems sizes, $L_1 , L_2 \gg \Lambda$,
we have
\begin{equation}
    W \approx \left(W_0-1\right)\ln(L_2/L_1)\simeq W_0\ln(L_2/L_1) \, .
\end{equation}
Thus, the work per particle over one cycle solely depends on $W_0$ and  its dependence on system size is reminiscent of the work done by a passive ideal gas.

\section{Efficiency} Finally, we define the efficiency 
of the work cycle as the ratio of the total work performed by the system composed of $N$ particles
\begin{equation}
 W^\text{tot}=NW
\end{equation}
to the total energy injected into the system,
\begin{equation}
    \eta=W^\text{tot}\big/(W^\text{tot}+W_\mathrm{irr}) \, ,
    \label{eq:eff}
\end{equation}
where $W_\mathrm{irr}$ accounts for the energy spent in dissipative processes. In the following we assume that the change in the value of P\'eclet number occurs with no additional dissipation, as it happens for example for light--controlled active colloids whose activity can be tuned by shading the light~\cite{Palacci2013,2017singh}.
Accordingly, $W_\mathrm{irr}$ is the sum of two contributions.
First,  we express the power dissipated by the particles due to their active motion as $W^\mathrm{pcl}_\mathrm{irr}=N\mathcal{P}\tau$,
where $\mathcal{P}$ is the mean power\footnote{Note that $\mathcal{P}$ is an average over free particles and particles trapped at the wall.} dissipated by a single particle during one work cycle of period $\tau$. 
We remark that for infinitely slow processes $\tau\rightarrow \infty$, $W^\mathrm{pcl}_\mathrm{irr}\rightarrow \infty$ and hence $\eta\rightarrow 0$. Therefore, active engines should be run at 
finite--time compression and expansion, a regime in which the additional dissipation due to friction forces acting when the container expands or compresses, $W^\mathrm{sys}_\mathrm{irr}$, should be accounted for~\cite{Kajelstrup_book}. 
Very generally, the power dissipated by friction forces can be expressed as
\begin{equation}
    \dot{W}=F\cdot v\,,
\end{equation}
where $F$ is the applied force and $v$ is the velocity. Within linear response theory~\cite{VandenBroeck2005} we have $F=\mathcal{W}\,v$, where $\mathcal{W}$ is the friction coefficient.
Thus, the dissipated power reads
\begin{equation}
   \dot{W}=\mathcal{W} v^2.
\end{equation}
In the case under study, we estimate the velocity via the change in the volume $v\simeq 2\Delta L/\tau$ during the cycling time $\tau$. Accordingly, the dissipated power is
\begin{equation}
    \dot{W}=4\mathcal{W} \frac{\Delta L^2}{\tau^2}\,,
\end{equation}
and hence the dissipated work during the cycle reads
\begin{equation}
   W_\text{irr}^\text{sys}=\dot{W} \tau=4\mathcal{W} \frac{\Delta L^2}{\tau}\,.
\end{equation}
We remark that, in leading order, $W_\text{irr}^\text{sys}$ does not depend on the parameters characterizing the active particles, in particular, it does not depend on $N$.
Assuming instantaneous changes of $\Pe$,
the velocity 
is 
$2(L_2-L_1)/\tau=2\Delta L/\tau$
and 
$W^\mathrm{sys}_\mathrm{irr}=4\mathcal{W} \Delta L^2/\tau
$~\cite{VandenBroeck2005,Schmiedl2007}, where $\mathcal{W}$ plays the role of an effective friction coefficient.
All in all we obtain
\begin{align}
W_\mathrm{irr}=W^\mathrm{pcl}_\mathrm{irr}+W^\mathrm{sys}_\mathrm{irr} = N \mathcal{P}\tau+ 4 \mathcal{W} \Delta L^2 \big/\tau
   \label{eq:diss}
\end{align}
where $\cal{P}$ and $\cal{W}$ are phenomenological parameters encoding, respectively, for the power dissipated by the active particles\footnote{For bacteria, $\mathcal{P}$ amounts to the metabolic cost of keeping the bacteria alive and swimming. For diffusiophoretic colloids, $\mathcal{P}$ amounts to the power dissipated in order to keep the imbalance in the bulk chemical potentials of the reactants and the reaction products. For light driven phoretic colloids, $\mathcal{P}$ amounts to the power dissipated (per particle) by the light source. For ABPs, under the assumption that all the energy ``consumed'' by the internal mechanism responsible for active displacement is transformed into motion, we have $\mathcal{P}_\text{ABP}=F\cdot v_0=\frac{k_BT}{D} v_0^2$, where $F$ is the effective driving force.}, which depends on the specific propulsion mechanism, and for the effective friction of the container.
Maximizing the efficiency with respect to $\tau$ amounts to minimizing $W_\mathrm{irr}$ 
which gives an optimal time of the cycle
\begin{equation}
 \tau_\text{opt}=(4\mathcal{W}\Delta L^2/N\mathcal{P})^{1/2}\,.
\end{equation}
Accordingly, even for quasi--static expansions, the efficiency $\eta$ is maximized for a \textit{finite} cycle time $\tau$ in stark contrast to the quasi--static limit $\tau\to \infty$ of thermodynamic engines for which $W^\mathrm{pcl}_\mathrm{irr}$ is replaced by the heat $Q$ that is independent of $\tau$. This is in agreement with the numerical results of Ref.~\cite{Fodor2020}.
Interestingly, $\tau_\text{opt}$ depends on the ratio between the dissipation in the system, $4\mathcal{W}\Delta L^2$ and that due to the active bath $N\mathcal{P}$. Hence, active engines exploiting many particles should be run at short cycling times, whereas the opposite holds for smaller particle numbers.

In the above calculations we have assumed that the particle density follows adiabatically the change in volume $\Delta L$. 
We recall that the relaxation time of the system can be estimated as
\begin{equation}
 \tau_\text{relax}\simeq \text{min}\left(\frac{\Delta L^2}{D},\frac{\Delta L}{v_\text{act}}\right)\,.
\end{equation}
Hence, the adiabatic assumption we made in deriving Eq.~\eqref{eq:work} is fulfilled when
$\tau_\text{relax}\ll \tau_\text{opt}$. For $\tau_\text{relax}=\Delta L/v_\text{act}$ that leads to 
\begin{equation}
N\ll \frac{4{\cal W} v_\text{act}^2}{\cal{P}}\,.
\label{eq:tau_best}
\end{equation}
Interestingly, Eq.~\eqref{eq:tau_best} shows that the maximum efficiency can be attained by small systems, whereas larger systems containing a larger number of particles $N$, will be suboptimal.

\section{Conclusions} 
Based on a one--dimensional model for run--and--tumble particles, we have derived an analytic expression for the
mechanical pressure ABPs exert on bounding walls. In the limit of large systems we reproduce the well--known
scaling of the bulk pressure with $\Pe^2$. In contrast, for either strongly confined micrometric ABPs or macroscopic 
ABPs with very large $\Pe$, the pressure scales with $\Pe$ and is no longer an intensive variable. We clearly 
rationalize the regimes where the different scalings are observed. Furthermore, two--dimensional Brownian dynamics
simulations of ABPs quantitatively agree with our analytic expression and thereby show its generality.\\
Our analytic 
formula for pressure allows to systematically explore basic features of confined active systems between
bulk-- and surface--driven behavior. For the recently introduced active engines we have calculated the work performed 
during one cycle in the quasi--static limit. It explicitly depends on the characteristic time scale $\Gamma^{-1}$, a feature 
that is absent in conventional thermodynamic engines. Furthermore, the efficiency is maximized at a \textit{finite} cycle 
rate due to the inherent dissipation, in clear contrast to thermodynamic engines where infinitely small rates avoid 
dissipation. Surprisingly, such an optimal cycling time is typical of ``small engines'' i.e., those engines exploiting a small number of active particles.

\newpage

\onecolumn

\pagestyle{plain}

\renewcommand{\thepage}{\arabic{page}}
\setcounter{page}{1}
\renewcommand\theequation{S\arabic{equation}}
\setcounter{equation}{0}

\renewcommand{\subsection}[1]{\vspace{0.5cm}\begin{center} \textbf{#1}\end{center}\vspace{0.3cm}}

\newcommand{\ee}{\operatorname{e}}


\begin{center}
\textbf{SUPPLEMENTAL MATERIAL\vspace{-0.2cm}}
\end{center}


\subsection{S1. Model}

\label{App:model}
In this section we analyze the dynamics of $N$ noninteracting active Brownian particles (ABPs) confined in $1D$ and suspended in an equilibrium thermal bath. The ABPs are characterized by an ``internal state'' that determines the preferred direction of motion. When the particles are in the ``up'' state, the active contribution $v_\text{act}$ to the overall velocity points to the right (parallel to $x$ axis), while in ``down'' state it points to the left (antiparallel to $x$ axis). The particles can randomly hop between the two states with the rate $\gamma$. Additionally, we assume that the particles experience a confining potential 
\begin{align}
    \beta U(z)=\begin{cases}
    f \frac{z-L}{L} & \text{for } z> L,\\
    0 & \text{for } -L \leqslant z \leqslant L,\\
    -f \frac{z+L}{L} & \text{for } z< -L,
    \end{cases}
\end{align}
such that when the center of mass of a particle is not in the region $-L\leqslant z \leqslant L$ there is a constant force pushing the particle back into the region. The choice of a piecewise linear potential allows us for further analytical insight.

In the following, instead of the number density of particles  $\tilde\rho\left(z\right)$, we use the dimensionless quantities which we get by rescaling the distance $x=z/L$ and by multiplying $\tilde\rho$ by the particle size $R$
\begin{equation}
    \rho\left(x\right)=R \tilde\rho\left(z=xL\right).
\end{equation}
The potential in rescaled variables is
\begin{align}
    \beta U(x)=\begin{cases}
    f (x-1) & \text{for }x> 1,\\
    0 & \text{for }-1 \leqslant x \leqslant 1,\\
    -f (x+1) & \text{for }x< -1,
    \end{cases}
\end{align}
and the normalization condition is
\begin{equation}
\int_{-\infty}^\infty \tilde\rho\left(z\right)\dd z=N\quad \mapsto \quad  \int_{-\infty}^\infty \rho\left(x\right)\dd x=N\frac{R}{L}. 
\label{eq:norm}
\end{equation}
The time evolution of the dimensionless probability distributions of particles in the ``up state'' $\Pu\left(x\right)$ and ``down state'' $\Pd\left(x\right)$ are governed by
\begin{subequations}\label{appB:de}
\begin{align}
\dot{\rho}_{\uparrow}\left(x\right) & =  -\derx J_{\uparrow}-\Gamma\frac{L^2}{R^2}\left[\Pu\left(x\right)-\Pd\left(x\right)\right],\\
\dot{\rho}_{\downarrow}\left(x\right) & =  -\derx J_{\downarrow}+\Gamma\frac{L^2}{R^2}\left[\Pu\left(x\right)-\Pd\left(x\right)\right],
\end{align}
\end{subequations}
where we have used the dimensionless time, measured in $L^2/D$ units. We have identified the fluxes as
\begin{subequations}
\begin{align}
J_{\uparrow}\left(x\right) & = -\left[\derx \Pu\left(x\right)-\frac{L}{R}\Pe \Pu\left(x\right)+\Pu\left(x\right)\beta \nabla U\left(x\right)\right],\\
J_{\downarrow}\left(x\right) & =  -\left[\derx \Pd\left(x\right)+\frac{L}{R}\Pe \Pd\left(x\right)+\Pd\left(x\right)\beta \nabla U\left(x\right)\right],
\end{align}
\end{subequations}
with
\begin{equation}\label{eq:def-Pe-Gamma}
\Pe  =  \frac{v_\text{act}R}{D}, \qquad \Gamma  =  \frac{\gamma R^{2}}{D}.
\end{equation}
By introducing the total (dimensionless) density $\Pp\left(x\right)$ and the density difference $\Pm\left(x\right)$
\begin{equation}
\Pp\left(x\right) =  \Pu\left(x\right)+\Pd\left(x\right), \qquad
\Pm\left(x\right) =  \Pu\left(x\right)-\Pd\left(x\right),
\end{equation}
we can rewrite Eqs.~\eqref{appB:de} as
\begin{subequations}
\begin{align}
\Ppdot\left(x\right) & =  -\derx \Jp\left(x\right),\label{appB:Ppdot}\\
\Pmdot\left(x\right) & =  -\derx \Jm\left(x\right)-2\Gamma \frac{L^2}{R^2}\, \Pm(x),\label{appB:Pmdot}
\end{align}
\end{subequations}
with 
\begin{subequations}
\begin{align}
\Jp\left(x\right) & =  -\derx \Pp\left(x\right)+\frac{L}{R}\Pe \Pm\left(x\right)-\Pp\left(x\right)\beta\,\derx U\left(x\right),\label{eq:J+}\\
\Jm\left(x\right) & =  -\derx \Pm\left(x\right)+\frac{L}{R}\Pe \Pp\left(x\right)-\Pm\left(x\right)\beta\,\derx U\left(x\right).\label{eq:J-}
\end{align}
\end{subequations}
We assume that there are no particles away from the box
\begin{equation}\label{eq:BC2}
\Pp\left(x=\pm\infty\right)  =  0, \qquad \Pm\left(x=\pm\infty\right) =  0,
\end{equation}
which additionally implies that for $x=\pm\infty$ there are no fluxes. In the steady state ($\Ppdot\left(x\right)=\Pmdot\left(x\right)=0$) Eqs.~\eqref{appB:Ppdot} and \eqref{eq:J+} give
\begin{equation}
    \Pm\left(x\right)=\frac{1}{\Pe}\frac{R}{L}\left[\derx \Pp\left(x\right)+\Pp\left(x\right)\beta\,\derx U\left(x\right)\right].
\end{equation}
Substituting the above expression in Eq.~\eqref{appB:Pmdot}, using Eq.~\eqref{eq:J-}, after some algebra we get the general equation
\begin{multline}\label{appB:general}
    \derx^3 \Pp+2\left(\derx^2 \Pp\right)\beta\,\derx U+\left(\derx \Pp\right)\left(3\beta\,\derx^2 U-\frac{L^2}{R^2}\Pe^2+\left(\beta\, \derx U\right)^2-2\Gamma\frac{L^2}{R^2}\right)\\
    +\Pp\left(\beta\,\derx^3 U+2\left(\beta\,\derx U\right)\beta\,\derx^2 U-2\Gamma\frac{L^2}{R^2} \beta\,\derx U\right)=0.
\end{multline}

For $-1<x<1$, where $U\left(x\right)=0$, Eq.~\eqref{appB:general} simplifies to
\begin{equation}
    \derx^3 \Pp_\mathrm{c}-\left(\derx \Pp_\mathrm{c}\right)\left(\Pe^2+2\Gamma\right)\frac{L^2}{R^2}=0,
\end{equation}
Since we require $\Pp\left(x\right)$ to have the symmetry $x\mapsto -x$, the solution of the above equation is 
\begin{equation}\label{eq:Pc}
    \Pp_\mathrm{c}(x)=A_\mathrm{c}\cosh(\kappac x)+B_\mathrm{c}, \qquad \kappa^2_\mathrm{c}=(\Pe^2+2\Gamma)\frac{L^2}{R^2},
\end{equation}
where $A_\mathrm{c}$ and $B_\mathrm{c}$ are, yet to be determined, constants.

For $x>1$, where $\beta\,\derx U=f$, Eq.~\eqref{appB:general} takes the form
\begin{equation}
    \derx^3 \Pp_\mathrm{r}+2 f (\derx^2 \Pp_\mathrm{r})+\left(f^2-\frac{L^2}{R^2}\Pe^2-2\Gamma\frac{L^2}{R^2}\right)\derx \Pp_\mathrm{r}-2\Gamma\frac{L^2}{R^2} f \Pp_\mathrm{r}=0,
\end{equation}
and the solution is
\begin{equation}\label{eq:Pr}
    \Pp_\mathrm{r}\left(x\right)=A_1 e^{\kappa_1 x}+A_2 e^{\kappa_2 x}+A_3 e^{\kappa_3 x},
\end{equation}
where $A_1$, $A_2$ and $A_3$ are constants, and $\kappa_{1}\geqslant \kappa_2 \geqslant \kappa_3$ are the roots of the polynomial
\begin{equation}\label{eq:kappa}
    \mathcal{Q}\left(\kappa\right)=\kappa^3+2 f \kappa^2+\kappa \left(f^2-\frac{L^2}{R^2}\Pe^2-2\Gamma\frac{L^2}{R^2}\right)-2\Gamma\frac{L^2}{R^2} f.
\end{equation}
Since $\mathcal{Q}(0)=-2\Gamma \frac{L^2}{R^2}f<0$ and $\mathcal{Q}\left(-f\right)=f \frac{L^2}{R^2}\Pe^2 >0$, the above polynomial has three different, real roots; $\kappa_1$ is positive and $\kappa_2, \kappa_3<0$. Using the Cardano's formula, after some algebra, we get
\begin{subequations}
\begin{align}\label{eq:k2k3}
    \kappa_1 &= -\frac{2 f}{3}+\frac{2}{3} \sqrt{f^2+3 \frac{L^2}{R^2} \left(2 \Gamma+\Pe^2\right)} \cos \frac{\theta}{3}, \\
    \kappa_2 &= -\frac{2 f}{3}  +\frac{1}{3} \sqrt{f^2+3 \frac{L^2}{R^2} \left(2 \Gamma+\Pe^2\right)} \left(\sqrt{3} \sin \frac{\theta}{3}-\cos \frac{\theta}{3} \right),\\
    \kappa_3 &= -\frac{2 f}{3} + \frac{1}{3} \sqrt{f^2+3 \frac{L^2}{R^2} \left(2 \Gamma+\Pe^2\right)} \left(\sqrt{3} \sin \frac{\theta}{3}+\cos \frac{\theta}{3} \right),
\end{align}
\end{subequations}
where the angle $\theta$ is given by
\begin{equation}\label{eq:theta}
\sin\theta=\frac{\left[\left(f^2+3 \frac{L^2}{R^2} \left(2 \Gamma+\Pe^2\right)\right)^3 -\left(f^3+9 f \frac{L^2}{R^2} \left(\Gamma-\Pe^2\right)\right)^2\right]^{1/2}}{\left[f^2+3\frac{L^2}{R^2}\left(\Pe^2+2\Gamma\right)\right]^{3/2}},\qquad \cos\theta=\frac{ f^3+9 f \frac{L^2}{R^2}\left(\Gamma-\Pe^2\right)}{\left[f^2+3\frac{L^2}{R^2}\left(\Pe^2+2\Gamma\right)\right]^{3/2}}.
\end{equation}
Due to the symmetry $x\mapsto -x$ there is no need to separately consider the case of $x<-1$.

In order to determine the integration constants we first impose the normalization condition (see Eq.~\eqref{eq:norm})
\begin{equation}
    \bar{\rho}=\frac{NR}{2L}=\int_0^\infty \Pp\left(x\right)\dd x  =  \frac{A_\text{c}}{\kappac}\sinh\kappac+B_\text{c} +\left.\frac{A_1}{\kappa_1}e^{\kappa_1 x}\right|^{\infty}_{1}+\left.\frac{A_2}{\kappa_2}e^{\kappa_2 x}\right|^{\infty}_{1}+\left.\frac{A_3}{\kappa_3}e^{\kappa_3 x}\right|^{\infty}_{1}.
    \label{eq:norm2}
\end{equation}
Since $\kappa_1>0$, the only way to satisfy the above condition is to require $A_1=0$. The resulting relation is
\begin{subequations}\label{appB:bnd_cnd}
\begin{align}
    \bar{\rho}&= \frac{A_\text{c}}{\kappac}\sinh\kappac+B_\text{c}-\frac{A_2}{\kappa_2}e^{\kappa_2 }-\frac{A_3}{\kappa_3}e^{\kappa_3}.\\
\intertext{Three more relations come from the requirement that $\Pp$, $\Pm$, $\Jp$, and $\Jm$ are continuous at $x=1$ (discontinuity of probability would lead to an infinite flux). The requirement of continuity of $\Pp$ gives}
\label{appB:bnd_cnd1}
    A_\text{c}\cosh{\kappac}+B_\text{c}&=A_2 e^{\kappa_2}+A_3 e^{\kappa_3},    
\intertext{the continuity of $\Pm$ gives}
     A_\text{c} \kappac \sinh\kappac&=A_2 \kappa_2 e^{\kappa_2}+A_3 \kappa_3 e^{\kappa_3}+f\left(A_2  e^{\kappa_2}+A_3  e^{\kappa_3}\right), 
\intertext{and the continuity of $\Jm$ gives (we have used Eq.~\eqref{appB:bnd_cnd1} to simplify the formula)}
    A_\text{c} \kappac^2 \cosh\kappac&=A_2 \kappa_2^2 e^{\kappa_2}+A_3 \kappa_3^2 e^{\kappa_3}+2f\left(A_2 \kappa_2  e^{\kappa_2}+A_3 \kappa_3  e^{\kappa_3}\right)+f^2\left(A_2 e^{\kappa_2}+A_3e^{\kappa_3}\right).
\end{align}
\end{subequations}
Since $\Jp\left(x\right)=0$, there is no equation coming from the requirement that $\Jp$ is continuous at $x=1$. The solution of the linear Eqs.~\eqref{appB:bnd_cnd} is 
\begin{subequations}\label{appB:sol}
\begin{align}
    A_\text{c}&=-\bar{\rho}\,\frac{\kappa_2 \kappa_3 \kappac \left(f+\kappa_2\right) \left(f+\kappa_3\right)}{G_2\cosh \kappac-G_1\sinh \kappac},\\
    B_\text{c}&=\bar{\rho}\,\frac{\kappa_2 \kappa_3 \kappac \left[ \left(\left(f+\kappa_2\right) \left(f+\kappa_3\right)+\kappac^2\right)\cosh \kappac- \left(2 f+\kappa_2+\kappa_3\right)\kappac \sinh \kappac\right]}{G_2\cosh \kappac-G_1\sinh \kappac},\\
    A_2&=\bar{\rho}\,\frac{e^{-\kappa_2}\kappa_2\kappa_3 \kappac^2\left(f+\kappa_3\right)\left[\left(f+\kappa_3\right)\sinh\kappac - \kappac \cosh\kappac\right]}{\left(\kappa_2-\kappa_3\right)\left(G_2\cosh \kappac-G_1\sinh \kappac\right)},\\
    A_3&=\bar{\rho}\,\frac{e^{-\kappa_3} \kappa_2\kappa_3 \kappac^2 \left(f+\kappa_2\right) \left[\left(f+\kappa_2\right) \sinh \kappac-\kappac \cosh \kappac\right]}{\left(\kappa_3-\kappa_2\right)\left(G_2\cosh \kappac-G_1\sinh \kappac \right)},\\
    \intertext{with}
    G_1&= \kappa_2 \kappa_3 \left(f+\kappa_2\right) \left(f+\kappa_3\right)- \kappac^2 \left[\left(f+\kappa_2\right)^2-\left(\kappa_2-1\right) \kappa_3 \left(2 f+\kappa_2\right)-\left(\kappa_2-1\right) \kappa_3^2\right],\\
    G_2&= \kappa_2 \kappa_3 \kappac \left(f+\kappa_2\right) \left(f+\kappa_3\right)- \kappac^3 \left(f-\kappa_2 \kappa_3+\kappa_2+\kappa_3\right).
\end{align}
\end{subequations}
Accordingly, the dimensionless mechanical pressure (measured in $k_\text{B} T/R$ units) of $N$ noninteracting particles is
\begin{multline}\label{appB:Pi1}
    \Pi=
    \int_1^\infty \beta\, \derx U\left(x\right) \Pp\left(x\right)\dd x=\int_1^\infty f \Pp\left(x\right)\dd x\\
    =
    \bar{\rho}f\kappac^2\,\frac{ \left[\kappa_2\left(f+\kappa_2\right)^2-\kappa_3\left(f+\kappa_3\right)^2\right]\sinh \kappac-\kappac \left[\kappa_2\left(f+\kappa_2\right)-\kappa_3\left(f+\kappa_3\right)\right] \cosh \kappac}{\left(\kappa_2-\kappa_3)(G_2\cosh \kappac-G_1\sinh \kappac\right)}.
\end{multline}

We are interested in deriving an expression for the pressure $\Pi$ in the limit $f\to \infty$, \textit{i.e.}, in the case in which the particle is confined in a box ($x\in [-1:1]$). From Eq.~\eqref{eq:kappa} for large $f$ we have
\begin{subequations}
\begin{align}
\kappa_1&=2 \Gamma \frac{L^2}{R^2}\,\frac{1}{f}+\mathrm{O}\left(\frac{1}{f^2}\right),\\
\kappa_2&=-f+\frac{L}{R} \Pe-\frac{L^2}{R^2} \Gamma \frac{1}{f}+\mathrm{O}\left(\frac{1}{f^2}\right),\\
\kappa_3&=-f-\frac{L}{R} \Pe-\frac{L^2}{R^2} \Gamma \frac{1}{f}+\mathrm{O}\left(\frac{1}{f^2}\right).
\end{align}
\end{subequations}
Using the above asymptotic expansion, after some algebra we get
\begin{subequations}
\begin{align}
    G_1&=-\Pe^2 \frac{L^2}{R^2} f^2+\mathrm{O}\left(f\right), & G_2&=2\Gamma \frac{L^2}{R^2}\kappac f^2+\mathrm{O}\left(f\right),\\
    A_\mathrm{c}&= \bar\rho\,\frac{\kappac \Pe^2}{\Pe^2\sinh \kappac+2\Gamma \kappac\cosh \kappac}+ \mathrm{O}\left(\frac{1}{f}\right), & B_\mathrm{c}&=\bar\rho\,\frac{2\Gamma \kappac \cosh\kappac}{\Pe^2\sinh \kappac+2\Gamma \kappac\cosh \kappac}+\mathrm{O}\left(\frac{1}{f}\right).
\end{align}
\end{subequations}
Finally, from Eq.~\eqref{appB:Pi1} we get
\begin{equation}
    \Pi=\bar\rho\,\frac{R^2}{L^2}\,\frac{\kappac^3 \cosh\kappac}{\Pe^2\sinh \kappac+2\Gamma \kappac\cosh \kappac}-\frac{1}{f}\,\bar\rho\,\frac{R^2}{L^2} \frac{\kappac^4\left(\Pe^2+2\Gamma \cosh^2\kappac\right)}{\left(\Pe^2\sinh \kappac+2\Gamma \kappac\cosh \kappac\right)^2}+\mathrm{O}\left(\frac{1}{f^2}\right),
\end{equation}
therefore
\begin{equation}
    \Pi_\infty=\lim_{f\to \infty} \Pi =
    \bar\rho\,\frac{R^2}{L^2}\,\frac{\kappac^3 \cosh\kappac}{\Pe^2\sinh \kappac+2\Gamma \kappac\cosh \kappac}.
    \label{eq:Pr-box}
\end{equation}
From Eqs.~\eqref{eq:Pc} and \eqref{appB:sol} for $-1\leqslant x\leqslant 1$
\begin{subequations}
\begin{align}
\Pp_\infty \left(x\right)&= \lim_{f\to \infty} \Pp\left(x\right)  =  \kappac\bar{\rho}\,\frac{\Pe^2 \cosh\left(\kappac x\right)+2\Gamma \cosh\kappac}{\Pe^2\sinh \kappac+2\Gamma \kappac\cosh \kappac},\\
\Pm_\infty \left(x\right)&= \lim_{f\to \infty} \Pm\left(x\right)  =  \kappac^2\bar{\rho}\,\frac{R}{L}\frac{\Pe \sinh\left(\kappac x\right)}{\Pe^2\sinh \kappac+2\Gamma \kappac\cosh \kappac}.
\end{align}
\end{subequations}
For $\left|x\right|>1$ we get $\Pp_\infty\left(x\right)=\Pm_\infty\left(x\right)=0$.

\subsection{S2. Pressure in different limiting cases}

In order to study the properties of the pressure in different limits, we expand the general result \eqref{eq:Pr-box} (Eq.~\eqref{eq:txt_Pr-box} of the main text) in the limit of large $\kappac$
\begin{equation}\label{eq:pexpansion1}
\Pi_\infty=\bar\rho\,\frac{R^2}{L^2}\,\frac{\kappac^3 }{\Pe^2\tanh \kappac+2\Gamma \kappac}=\bar\rho\,\frac{R^2}{L^2}\,\frac{\kappac^3 }{\Pe^2+2\Gamma \kappac}+\mathrm{O}\left(\ee^{-2\kappac}\right),
\end{equation}
which proves Eq.~\eqref{eq:txt_Pr-box-approx} of the main text. We note that the corrections to the above equation decay exponentially fast upon increasing $\kappac$, therefore this approximation is reliable even for moderate values of $\kappac$. For the discussion of physical arguments justifying this limit, see the main text. Further expansion of \eqref{eq:pexpansion1} for large $\kappac$ gives
\begin{equation}\label{eq:pexpansion2}
\Pi_\infty\simeq \bar\rho\,\frac{R^2}{L^2}\,\frac{\kappac^2 }{2\Gamma}\frac{1}{1+\frac{\Pe^2}{2\Gamma \kappac}}=\bar\rho \left(1+\frac{\Pe^2}{2\Gamma }\right)\frac{1}{1+\frac{\Pe^2}{2\Gamma \kappac}}=\bar\rho \left(1+\frac{\Pe^2}{2\Gamma }\right)+\mathrm{O}\left(\frac{\Pe^2}{2\Gamma \kappac}\right).
\end{equation}
In the limiting case $\Pe\ll\Pesmall=\sqrt{2\Gamma}$, the correction term in the above equation is very small. Therefore for $\Pe\ll\Pesmall$ the approximation~\eqref{eq:pexpansion2} is again reliable even for moderate values of $\kappac$. This proves Eq.~\eqref{eq:pi_small} of the main text.

Finally, we study the pressure in the limit $\Pe\gg\Pesmall$. In this limit
\begin{equation}\label{eq:kappac:limit}
\kappac=\sqrt{\Pe^2+2\Gamma}\frac{L}{R}\simeq \Pe \frac{L}{R}.
\end{equation}
Using Eq.~\eqref{eq:kappac:limit}, Eq.~\eqref{eq:pexpansion1} can be simplified to
\begin{equation}
\Pi_\infty\simeq\bar\rho\,\frac{R^2}{L^2}\,\frac{\kappac^3 }{\Pe^2+2\Gamma \kappac}=\bar\rho \frac{L}{R}\frac{\Pe^2}{\Pe+2\Gamma \frac{L}{R}}=\bar\rho \frac{L}{R}\frac{\Pe^2}{\Pe+\Pelarge},
\end{equation}
where $\Pelarge=2\Gamma L/R$. This proves that
\begin{subequations}\label{eq:pexpansion34}
\begin{align}
\label{eq:pexpansion3}\Pi_\infty&\simeq\bar\rho \frac{L}{R}\,\frac{\Pe^2}{\Pelarge}=\bar\rho\frac{\Pe^2}{2\Gamma} && \hspace{-3cm}\text{for } \Pe\ll\Pelarge \text{ and }\Pe\gg\Pesmall,\\
\label{eq:pexpansion4}\Pi_\infty&\simeq\bar\rho \frac{L}{R}\, \Pe && \hspace{-3cm}\text{for  }\Pe\gg\Pelarge \text{ and }\Pe\gg\Pesmall,
\end{align}
\end{subequations}
which is equivalent to Eq.~\eqref{eq:pi_large} of the main text. 

In the above derivation we have not used the fact that $\kappac$ is large. Therefore, like Eq.~\eqref{eq:pexpansion1}, the result is reliable even for moderate values of $\kappac$.

We note that, there is a range of parameters of the system for which $\Pesmall$ and $\Pelarge$ are of the same order, or even $\Pesmall>\Pelarge$. In that case, the regime where $\Pi_\infty \propto \Pe^2$ (\eqref{eq:pexpansion3}) is not present and, upon increasing $\Pe$, the behavior of the pressure changes directly from independent of $\Pe$ (Eq.~\eqref{eq:pexpansion2}) to linear in $\Pe$ (Eq.~\eqref{eq:pexpansion4}).

\subsection{S3. Effective Temperature}

	Eq.~\eqref{eq:pi_small} of the main text (and Eq.~\eqref{eq:pexpansion2}) suggests to introduce the effective temperature
\begin{align}
    T_\text{eff}= T\left(1+\frac{\Pe^2}{2\Gamma}\right),
    \label{eq:T_eff}
\end{align}
as it is done in Refs.~\cite{Howse2007SM,Palacci2010SM} (Refs.~\cite{Howse2007} and \cite{Palacci2010} of the main text). However, we remark that in our case this is only possible in the regime $\Pe\ll \Pesmall$, for which Eq.~\eqref{eq:pi_small} of the main text holds. As shown in Eq.~\eqref{eq:pi_large} of the main text (see~Eq.~\eqref{eq:pexpansion34}), in the regime $\Pe \gg \Pelarge$ we cannot introduce $T_\text{eff}$ anymore since we are not in the limit of large tumbling rate $\Gamma$, where $\Gamma$ can be eliminated adiabatically.


\subsection{S4. Comparison with Ref.~\cite{Tailleur2015SM}}\label{app:cmp_cates}

In order to compare our results with Ref.~\cite{Tailleur2015SM} (Ref.~\cite{Tailleur2015} of the main text) we calculate
\begin{align}
    \frac{\Pi_\infty}{\Pp_\infty(x=0)}=\frac{(\Pe^2+2\Gamma)\cosh(\kappac)}{\Pe^2+2\Gamma \cosh(\kappac)},
\end{align}
which for $\kappac \gg 1$ reduces to
\begin{align}
    \frac{\Pi_\infty}{\Pp_\infty(x=0)}=\frac{\Pe^2}{2\Gamma}+1, 
\end{align}
the result reported in Ref.~\cite{Tailleur2015SM}. Note that $\kappac=\frac{L}{R}\sqrt{\Pe^2+2\Gamma} \gg 1$ can be attained for systems whose size is 
\begin{align}
    L\gg \frac{R}{\sqrt{\Pe^2+2\Gamma}}
\end{align}
The discrepancy between our formula (Eq.~\eqref{eq:txt_Pr-box} of the main text and Eq.~\eqref{eq:Pr-box}) and Ref.~\cite{Tailleur2015SM} is relevant for weakly active systems, \textit{i.e.}, for  $\Pe \lesssim 1$ and $\Gamma \lesssim 1$.
We note that, thanks to our approach, we can compute the full value of $\Pi_\infty$ on the top of its ``deviation'' from the ideal gas law,  $\Pi_\infty/\Pp_\infty(x=0)$.

\subsection{S5. Compressibility}
Here we derive the expression for the compressibility
\begin{equation}\label{eq:compr}
    H_\infty= -\frac{1}{L}\left(\frac{\partial \Pi_\infty}{ \partial L}\right)^{\!-1}
    =\frac{1}{\bar{\rho}}\left(\frac{L}{R}\right)^2\frac{\left(\Pe^2 \sinh \kappac+2\Gamma \kappac \cosh \kappac\right)^2}{\kappac^4 \left(2\Gamma \cosh \kappac+\Pe^2\right)} =\frac{1}{\Pi_\infty}\frac{\left(2\Gamma \kappac\cosh\kappac+\Pe^2\sinh\kappac\right)\cosh\kappac}{\kappac\left(\Pe^2+2\Gamma \cosh^2\kappac\right)}.
\end{equation}
Notice that in the limit $\kappac\gg 1$ the last expression reduces to
\begin{align}
  H_\infty&=  \frac{1}{\Pi_\infty}\frac{2\Gamma \kappac+\Pe^2}{2\Gamma \kappac}.
\end{align}

\subsection{S6. Numerical simulations}

In this section we describe the numerical simulations of Brownian dynamics of ABPs we have preformed to support the results reported in this manuscript. The comparison with exact formulae is presented in Fig.~\ref{fig:pressure_model}(b) of the main text.

We consider a two--dimensional system of $N$ particles confined in a square box of size $2L\times 2L$. Each particle is described by a position vector $\mathbf{x}_i\left(t\right)$ and an angle $\theta_i\left(t\right)$ defining the orientation, where $i=1,2,\ldots, N$ labels the particles and $t$ is the time. The equations of motion are
\begin{subequations}\label{app:EM}
\begin{align}
   \mathbf{x}_i\left(t+\dd t\right)&=\mathbf{x}_i\left(t\right)+\mathbf{w}_i\left(t\right)\dd t=\mathbf{x}_i\left(t\right)+\left[\mathbf{v}_i\left(t\right)+\sqrt{\frac{2D}{\dd t}} \mathbfcal{R}_{i}\left(t\right)\right]\dd t,\\
  \label{app::EMrad} \theta\left(t+\dd t\right)&=\theta\left(t\right)+\sqrt{\frac{2D\Gamma}{R^2\dd t}} \chi\left(t\right)\dd t,
\end{align}
\end{subequations}
where $\dd t$ is the time step, $\mathbf{w}_i$ is the average velocity of the particle over the time $\dd t$, $\mathbf{v}_i=\left[ v_\text{act} \cos\theta_i\left(t\right), v_\text{act} \sin\theta_i\left(t\right)\right]$ is the active velocity, and $\mathbfcal{R}_{i}$ and $\chi$ denote independent random variables with normal distribution with zero mean value and unit variance that model the random noise. We note that in Eq.~\eqref{app::EMrad} we have assumed the rotational diffusion constant $D_\mathrm{r}=D \Gamma/R^2=\gamma$ to be equal to the tumbling rate of the particles in 1D model discussed in the main text. This simple assumption is enough to observe the agreement between 1D model and numerical simulations in 2D shown in Fig.~\ref{fig:pressure_model}(b) of the main text.

In the simulation we assume periodic boundary conditions in vertical direction. If the equations of motion Eq.~\eqref{app:EM} move the $i$-th particle beyond the left or right wall, we assume that $\mathbf{x}_i\left(t+\dd t\right)=\mathbf{x}_i\left(t\right)$ (resetting the position of the particle) and over the time $\dd t$ the average force exerted on the wall by the particle is
\begin{equation}
    F_i=\frac{k_B T}{D} \mathbf{w_i}\cdot \mathbf{n},
\end{equation}
where $D/\left(k_B T\right)$ is the mobility given by the Einstein relation, and $\mathbf{n}$ is the unit normal vector of the wall ($\mathbf{n}=\left[1,0\right]$ for the right wall and $\mathbf{n}=\left[-1,0\right]$ for the left wall). During the simulation, we have been calculating sums of all the forces exerted by the particles in a given time step, we average this quantity over all time steps, and divide it by the length of the wall to obtain the pressure.

Since the pressure presented in Fig.~\ref{fig:pressure_model} of the main text is normalized by the pressure of ideal gas, the exact value of $T$ (and the height of the system which is assumed to be $2L$) is not relevant. Moreover, changing $D$ and $R$ (with $L/R$ fixed) is equivalent to rescaling of $\dd t$. Therefore, for the simulation we have taken $k_B T=D=R=1$. For several values of the parameters we have checked that when $\dd t\lesssim 0.01$, the pressure 
is de facto independent on the exact value of $\dd t$; therefore we have assumed $\dd t=0.01$.

For each value of $\Pe$, $\Gamma$ and $L/R$ we have prepared the initial configuration by placing $N=200$ particles in the box randomly and with random orientation. Then, the simulation was run for $10^6$ time steps in order to relax the initial condition. 
Finally, for $10^7$ time steps the pressure was measured. The simulation for each set of parameters has been repeated 100 times and the average pressure and its standard deviation calculated. In each case the standard deviation of the pressure is much smaller that the size of points in the plot.



\end{document}